\documentclass[prx,preprint,preprintnumbers]{revtex4}
\usepackage{graphicx} % Include figure files; subfigure
\usepackage{dcolumn} % Align table columns on decimal point
\usepackage{bm} % bold math
\usepackage{amsmath}

\usepackage[dvips]{color}

\usepackage{epsfig}
\newcommand{\nn}{\nonumber}

\begin{document}
\title{Metal nanospheres under intense continuous wave illumination - a unique case of non-perturbative nonlinear nanophotonics}

% Let Ieng Wai, SWC , Said, Ohad read this...

% version 18 - with comments from Ioseph...; PR format.

\author{I. Gurwich, Y. Sivan}\email{sivanyon@bgu.ac.il}
\affiliation{Unit of Electro-optics Engineering, Faculty of Engineering Sciences, Ben-Gurion University of the Negev, Israel}

\begin{abstract}
  We show that the standard perturbative (i.e., cubic) description of the thermal nonlinear response of small metal nanospheres to intense continuous wave illumination is insufficient already beyond temperature rises of a few tens of degrees. In some cases, a cubic-quintic nonlinear response is sufficient to describe accurately the intensity dependence of the temperature, permittivity and field, while in other cases, a full non-perturbative description is required. We further analyze the relative importance of the various contributions to the thermal nonlinearity, identify a qualitative difference between Au and Ag, and show that the thermo-optical nonlinearity of the host typically plays a minor role, but its thermal conductivity is important. % We exemplify the intricate interplay between the various effects in a variety of realistic examples.
\end{abstract}
\maketitle

\section{Introduction}

Nonlinear optical effects are usually considered weak. For example, typical second-order nonlinear processes such as second harmonic generation have efficiencies of a few percent for nonlinear crystals a few mm in length~\cite{Boyd-book}. Third-order nonlinear effects are usually weaker and occur on even longer distances, see for example, self-phase modulation in optical fibers~\cite{Agrawal-book} or even in highly nonlinear semiconductor waveguides~\cite{Agrawal_Painter_review_OE_2007}. Accordingly, the vast majority of theoretical studies of nonlinear optical effects are performed within the framework of a perturbative description, i.e., involving, at most, a third-order (equivalently, cubic) nonlinearity.

Ultrafast nonlinearities are particularly weak. However, even the much stronger slow nonlinearities are limited to modest permittivity changes due to saturation effects (see~\cite{Khurgin_chi3} and references therein), phase transitions (e.g., in liquid crystals), or damage.

Here, we study a case, unique, to the best of our knowledge, whereby the perturbative description is insufficient, sometimes even requiring to go beyond a cubic-quintic nonlinearity to a full non-perturbative description. In particular, we consider a continuous wave (CW) illumination of a few nm metal sphere. We will focus on the {\em thermal} effect, known to be among the strongest sources of optical nonlinearity, with the dual aim of demonstrating the failure of the perturbative (cubic) description and identifying the importance of the various underlying physical effects. While doing so, we will elucidate the quantitative and qualitative differences between the strength of the thermal nonlinearity in the ultrafast and CW cases. This is important because while the former is well-studied, the latter, i.e., the CW case, has been practically ignored.

In that sense, this study was motivated by recent measurements of the extremely strong scattering of intense CW light from small metal particles~\cite{plasmonic-SAX-ACS_phot,plasmonic-SAX-PRL,plasmonic-SAX-OE,plasmonic-SAX-rods-Ag,SUSI}, which, although being conceptually simple, are first of their kind. Further, this study was made possible by recent ellipsometry measurements of metal permittivities at elevated temperatures~\cite{Wilson_deps_dT,Indian_Ag_ellipsometry_2014,Shalaev_ellipsometry_gold,PT_Shen_ellipsometry_gold}. Thus, although we present here rather simple theoretical calculations, we emphasize that the approach is quite realistic. Indeed, it was already shown in~\cite{Sivan-Chu-high-T-nl-plasmonics} that the thermal effect reproduces {\em qualitatively} the experimental results of~\cite{plasmonic-SAX-ACS_phot,plasmonic-SAX-PRL,plasmonic-SAX-OE,plasmonic-SAX-rods-Ag,SUSI}. In the current article, we go beyond~\cite{Sivan-Chu-high-T-nl-plasmonics} and study this problem systematically and quantitatively, classifying the importance of the various contributing physical effects using numerical simulations and an approximate analysis. We identify qualitative differences between Au and Ag spheres, as well as between various hosts. This allows us to show that unless the host undergoes a phase transition (e.g., boiling), % or other effects like stress/strain
it has a relatively smaller role compared to the metal. Overall, all cases require a description beyond the standard perturbative description of the nonlinear response. In some cases, a cubic-quintic nonlinear response can capture the intensity dependence of the temperature, permittivity and field, while in other cases, a full non-perturbative description is required.

%We start by deriving an analytical expression that shows the scaling of the thermal effect in small spheres with particle size, intensity and thermo-derivatives. This expression should then be generalized to other structures which require more sophisticated numerical simulations.

This study provides an important step towards reaching a quantitative understanding of the nonlinear scattering observed in~\cite{plasmonic-SAX-ACS_phot,plasmonic-SAX-PRL,plasmonic-SAX-OE,plasmonic-SAX-rods-Ag,SUSI} (i.e., matching theoretical predictions to the experimental data). It also allows us to predict so far unobserved behaviour, so that if observed experimentally, further support to the claims of the dominance of the thermal effects will be obtained.

The paper is organized as follows. In Section~\ref{sec:assumptions}, we describe the basic assumptions of the model, and in Section~\ref{sec:model} we develop the model equations for the temperature, permittivity and field within the nanosphere. We then proceed by several numerical examples (Section~\ref{sec:examples}) and complement the numerical results with an approximate analysis (Section~\ref{sec:analysis}). Section~\ref{sec:discussion} provides a discussion of the results and an outlook.

\section{Model assumptions}\label{sec:assumptions}
In what follows, we denote the quantities within the metal nanosphere (host) with a subscript $m$ ($h$), and label the ambient properties by a $0$ superscript.

We consider spheres which are sufficiently small so that the field and temperature within the metal nanosphere ($E_m$ and $T_m$, respectively) are approximated as being uniform. The former assumption, known as the quasi-static approximation, is justified for particles sufficiently small with respect to the wavelength; the latter is valid when the host has a relatively lower thermal conductivity compared with the sphere (i.e., $\kappa_h \ll \kappa_m$). % ; this condition is satisfied typically for any metal or semiconductor sphere in a dielectric host\footnote{A semiconductor host will be, thus, useful for fast switching applications, see~\cite{Khurgin-diffusive-switching}. }.

We assume that the nanospheres exhibit only linear (one-photon) absorption. While we do not expect multi-photon absorption to yield anything more than a quantitative change to the results shown below, it is also worth noting that the absorption cross-sections of multi-photon processes are difficult to distinguish from the many other effects (e.g., the thermal effect, free-carrier generation, stress/strain etc.), and are in general not well characterized, hence, difficult to quantify at this stage.

Further, we assume a linear dependence of the permittivity on the temperature, namely,
\begin{equation}\label{eq:eps_T_linear}
\epsilon_m(T) = \epsilon_{m,0}' + \frac{d\epsilon'_m}{dT}\bigg|_{T_h^0} (T - T_h^0) + i \left(\epsilon_{m,0}'' + \frac{d\epsilon''_m}{dT}\bigg|_{T_h^0} (T - T_h^0)\right),
\end{equation}
where $T_h^0$ is the temperature of the host away from the nanosphere. This was shown to be the case at least up to several hundreds of degrees in recent ellipsometry measurements of Au~\cite{Shalaev_ellipsometry_gold,PT_Shen_ellipsometry_gold} thin films; the data in these studies was remarkably similar %~[Spector, Gurwich, Sivan, A first principles model for high temperature permittivity of noble metals, in preparation.]
and was in excellent agreement with an earlier study~\cite{Wilson_deps_dT} performed over a narrower temperature range. Similar behaviour was also seen for a Ag film~\cite{Indian_Ag_ellipsometry_2014}, which describes probably the most detailed study, to date, of the temperature dependence of the Ag permittivity (in terms of wavelength and temperature range); it also includes an attempt to eliminate the effects of roughness, and is complemented with a first-principles calculation of the permittivity. It should be noted, however, that in contrast to the case of Au, the Ag results of~\cite{Indian_Ag_ellipsometry_2014} were not always in agreement with earlier studies; the authors of~\cite{Indian_Ag_ellipsometry_2014} included a detailed comparison and a discussion of the possible reasons for these differences. The most probable reason is the relatively large grain size, and the potential modification of the metal volume morphology at high temperatures, during and after the measurement themselves, see corresponding discussion in~\cite{Shalaev_ellipsometry_gold,PT_Shen_ellipsometry_gold}.

Similarly, we assume for the host media that $\epsilon_h = \epsilon_{h,0}' + d\epsilon'_h/dT|_{T_h^0} (T - T_h^0)$. Finally, we assume that the thermal conductivity also varies linearly with temperature, i.e., $\kappa_h = \kappa_{h,0} + d\kappa_h/dT|_{T_h^0} (T - T_h^0)$. Under the current conditions, the thermal conductivity can also account for the temperature dependence of the Kapitza resistance via a slight modification of the solution adopted here (see Eq.~(\ref{eq:T_self-consistent}) below)~\cite{Sivan-Chu-high-T-nl-plasmonics,Baffou_pulsed_heat_eq_with_Kapitza}.

For brevity, in what follows, we denote $B_m \equiv d\epsilon_m/dT|_{T_h^0} = B_m' + i B_m''$, $B_h' = d\epsilon'_h/dT|_{T_h^0}$ and $B_{\kappa,h} \equiv d\kappa_h/dT|_{T_h^0}$. For simplicity, we consider hosts which exhibit negligible absorption ($B_h'' = 0$), however, any non-zero contribution will cause only a quantitative difference.

% \footnote{\bf In the ultrafast regime, in~\cite{Opt_Limiting_NUS_2008} it was shown that 2 photon processes in rods and coupled spheres (mimicking a rod) are in fact two consequent one photon processes..; }

% SWC showed no metal luminescence above $10^5 W/cm^2$ CW for visible illumination~\cite{plasmonic-SAX-ACS_phot} - but this has nothing to do with MPA...

% Finally, we focus on the optical regime, in the spectral vicinity of the plasmon resonances, and since (as noted in~\cite{Sivan-Chu-high-T-nl-plasmonics}) the IR regime is not so interesting in the context of nonlinear optical response due to the lack of intense sources\footnote{There seems to be no experimental data for the permittivity at high temperatures in this regime, nor experimental studies; {\bf there is a model for Au (wrong in~\cite{high-T-plasmonics-Italians-1,high-T-plasmonics-Italians}).} }.

\section{Model} \label{sec:model}

We showed in~\cite{Sivan-Chu-high-T-nl-plasmonics} that the temperature inside a few nm sphere under intense monochromatic illumination is uniform, and given by
\begin{eqnarray}\label{eq:T_self-consistent}
\Delta T \equiv T_m - T_h^0 = \frac{a^2}{3 \kappa_h(T)} p_{abs}(T), \quad \quad p_{abs}(T) \equiv \frac{\omega}{2} \epsilon_0 \epsilon''_m(T) \left|\vec{E}_m\right|^2,
\end{eqnarray}
where $a$ is the sphere radius, $\omega$ is the angular frequency of the incoming photons, $\epsilon_0$ is the vacuum permittivity, and where
\begin{equation}\label{eq:E_np}
\vec{E}_m = \frac{3 \epsilon_h(T)}{\epsilon'_{tot}(T) + i \epsilon''_m(T)} \vec{E}_{inc}, \quad \quad
\end{equation}
is the field inside the nanosphere, $\vec{E}_{inc}$ is the incident field amplitude, and~$\epsilon'_{tot}(T) \equiv 2 \epsilon_h(T) + \epsilon_m'(T)$.

With the assumptions above, Eqs.~(\ref{eq:T_self-consistent})-(\ref{eq:E_np}) reduce to a 4-th order polynomial equation in $\Delta T$, namely,
\begin{eqnarray}\label{eq:DT--polyn-norm}
&& \underbrace{\frac{B_{\kappa,h}}{\kappa_{h,0}} \frac{{B'_{tot}}^2 + {B_m''}^2}{{\epsilon_{m,0}''}^2}}_{a_4} \Delta T^4 + \underbrace{\left[\frac{{B'_{tot}}^2 + {B_m''}^2}{{\epsilon_{m,0}''}^2} + 2 \frac{B_{\kappa,h}}{\kappa_{h,0}} \frac{B'_{tot} \epsilon_{tot}'B_m'' \epsilon''_{m,0}}{{\epsilon_{m,0}''}^2} - \Delta T_{lin}^{on} \frac{B_h'^2}{\epsilon_{h,0}'^2} \frac{B_m''}{\epsilon_{m,0}''}\right]}_{a_3} \Delta T^3 \nn \\
&& + \underbrace{\left[2 \frac{B'_{tot} \epsilon'_{tot} + B_m'' \epsilon''_{m,0}}{{\epsilon_{m,0}''}^2} + \frac{B_{\kappa,h}}{\kappa_{h,0}} \left(1 + \frac{{\epsilon'_{tot}}^2}{{\epsilon_{m,0}''}^2}\right) - \Delta T_{lin}^{on} \left(2\frac{B'_h}{\epsilon_{h,0}'} \frac{B_m''}{\epsilon_{m,0}''} + \frac{{B'_h}^2}{\epsilon_{h,0}'^2}\right)\right]}_{a_2} \Delta T^2 \nn \\
&& + \underbrace{\left[\left(1 + \frac{{\epsilon'_{tot}}^2}{{\epsilon_{m,0}''}^2}\right) - \Delta T_{lin}^{on} \left(\frac{B_m''}{\epsilon_{m,0}''} + 2 \frac{B'_h}{\epsilon_{h,0}'}\right)\right]}_{a_1} \Delta T = \Delta T_{lin}^{on},
\end{eqnarray}
where $B'_{tot} \equiv B_m' + 2 B_h'$ and
\begin{equation}\label{eq:DT_lin_on}
\Delta T_{lin}^{on} \equiv \frac{3 a^2 \omega \epsilon_0 \epsilon_{h,0}'^2 |\vec{E}_{inc}|^2}{2 \kappa_{h,0} \epsilon''_{m,0}} = \frac{3 a^2 \omega n_{h,0}^3 I_{inc}}{4 c \kappa_{h,0} \epsilon''_{m,0}} > 0,
\end{equation}
is the temperature rise in the on-resonance linear case~\footnote{Note that $\Delta T_{lin}^{on}$ scales with $a^2$ because it is the ratio of the absorption within the volume, $\sigma_{abs} \sim a^3$, and the rate of heat transfer to the host, given by $(\kappa_{h,0}/a)$ times the surface area $a^2$ through which heat leaks to the host. } (see below) and where $I_{inc} = 2 \sqrt{\epsilon_{h,0}'} |\vec{E}_{inc}|^2 / Z_0$ is the incident intensity. More generally, $\Delta T_{lin}^{on}$ is proportional to $P_{abs}% \equiv \sigma_{abs} I_{inc}
$, the density of power absorbed within the sphere, namely, $\Delta T_{lin}^{on} = P_{abs}/4 \pi \kappa_{h,0} a$~\cite{thermo-plasmonics-review}. Eq.~(\ref{eq:DT_lin_on}) shows that under the quasistatic approximation used here, a change of the nanosphere radius or thermal conductivity of the host is equivalent to a rescaling of the incident intensity.

We note that all the coefficients in Eq.~(\ref{eq:DT--polyn-norm}) involve only the ratios of a thermo-derivatives of the respective parts of the permittivities and the permittivities themselves ($\epsilon_{m,0}''$ or $\epsilon_{h,0}'$). % These ratios will appear in the analysis below.

It is clear that one can distinguish between two generic scenarios - when the illumination is on-resonance ($\epsilon'_{tot} \ll \epsilon_{m,0}''$) or off-resonance (otherwise%$\epsilon'_{tot} \ge \epsilon_{m,0}''$
)% ; typically, the resonant spectral band is several tens of nm wide
. We showed in~\cite{Sivan-Chu-high-T-nl-plasmonics} that there is a qualitative difference between these two cases, something which can, in fact, be observed from the solution of the linear case (i.e., the solution of Eq.~(\ref{eq:DT--polyn-norm}) where all the thermo-derivatives vanish),
\begin{equation}\label{eq:DT_lin}
\Delta T_{lin} = \frac{{\epsilon''_{m,0}}^2}{{\epsilon''_{m,0}}^2 + {\epsilon'_{tot}}^2} \Delta T_{lin}^{on}. %  = \frac{\epsilon''_{m,0}}{{\epsilon''_{m,0}}^2 + {\epsilon'_{tot}}^2} \frac{3 a^2 \omega n_{h,0}^3}{4 c  \kappa_{h,0}} I_{inc}.
\end{equation}
Indeed, Eq.~(\ref{eq:DT_lin_on}) shows that in the off-resonance case, the temperature rise is higher for increasing $\epsilon''_{m,0}$, while the opposite is true for the on-resonance case.

% {\bf we already see that one cannot associate the metal with an intrinsic thermal nonlinear coefficient in the CW case because ... but is this not general? }

With the solution for the temperature in hand, the permittivity is given by Eq.~(\ref{eq:eps_T_linear}) and the field within the nanosphere by Eq.~(\ref{eq:E_np}). Specifically,
\begin{eqnarray}\label{eq:E_np_explicit}
\vec{E}_m\left(\Delta T(E_{inc})\right) &=& \frac{3 (\epsilon_{h,0}' + B_h' \Delta T)}{\epsilon_{tot}' + B_{tot}' \Delta T + i \epsilon_{m,0}'' + i B_m'' \Delta T} \vec{E}_{inc} \nn \\
&=& \frac{3 \epsilon_{h,0}'}{\epsilon_{tot}' + i \epsilon_{m,0}''} \left(\frac{1 + \frac{B_h'}{\epsilon_{h,0}'} \Delta T}{1 + \frac{B_{tot}' + i B_m'' }{\epsilon_{tot}' + i \epsilon_{m,0}''}\Delta T}\right) \vec{E}_{inc}, % \nn \\
\end{eqnarray}
where~\cite{West_alternat_plas_rev,Hache-cubic-metal-nlty}
\begin{equation}\label{eq:f}
f \equiv 3 \epsilon_{h,0}'/(\epsilon_{tot}' + i \epsilon_{m,0}''),
\end{equation}
is the usual Figure of Merit for field enhancement inside a nanosphere~\cite{West_alternat_plas_rev}, sometimes referred to as the Faraday number~\cite{NFE-thermo-plas-FOM}. At resonance, this reduces to
\begin{eqnarray}\label{eq:E_np_explicit_res}
\vec{E}_m\left(\Delta T(E_{inc})\right) &=& \frac{3 \epsilon_{h,0}'}{i \epsilon_{m,0}''} \frac{1 + \frac{B_h'}{\epsilon_{h,0}'} \Delta T}{1 + \left(\frac{B_m''}{\epsilon_{m,0}''} - i \frac{B_{tot}'}{\epsilon_{m,0}''}\right) \Delta T} \vec{E}_{inc} \nn \\
&\cong& \frac{3 \epsilon_{h,0}'}{i \epsilon_{m,0}''} \left(1 + \frac{B_h'}{\epsilon_{h,0}'} \Delta T\right) \left(1 - \left(\frac{B_m''}{\epsilon_{m,0}''} - i \frac{B_{tot}'}{\epsilon_{m,0}''} \right) \Delta T\right) \vec{E}_{inc} \nn \\
&\cong& - i f^{(res)} \left[1 - \left(\frac{B_m''}{\epsilon_{m,0}''} - \frac{B_h'}{\epsilon_{h,0}'} - i \frac{B_{tot}'}{\epsilon_{m,0}''}\right) \Delta T\right] \vec{E}_{inc} \nn \\
&=& \left[1 - \left(\frac{B_m''}{\epsilon_{m,0}''} - \frac{B_h'}{\epsilon_{h,0}'} - i \frac{B_{tot}'}{\epsilon_{m,0}''}\right) \Delta T\right] \vec{E}_m \left(T = T_h^0\right),
\end{eqnarray}
where $f^{(res)} \equiv 3 \epsilon_{h,0}'/\epsilon_{m,0}''$. Note that we expanded the denominator to first order accuracy in $\Delta T$, the accuracy level of Eq.~(\ref{eq:eps_T_linear}).

% {\bf in order to understand the scattered field, we need to know $\Delta \epsilon_{tot}';$ vs. $\epsilon_{m,0}''$... }

If the intensity is low, then $\Delta T \to \Delta T_{lin}$. In this case, Eqs.~(\ref{eq:DT_lin_on})-(\ref{eq:DT_lin}) provide the perturbative description of the thermal nonlinearity of metals (as a cubic nonlinearity), where $\chi_m^{(3)} |\vec{E}_m|^2 \sim B_m \Delta T_{lin}$~\cite{Hache-cubic-metal-nlty}. Any deviation from this solution, specifically, employing higher intensities, or due to a deviation from the linear dependence of the coefficients on the parameters will give rise to a deviation from the standard perturbative description of a cubic nonlinear response. While this may not be a new or surprising observation, unlike the well-studied ultrafast nonlinearity, the CW nonlinearity of metals in general, and of a single metal nanosphere, in particular, was not studied systematically at all. Accordingly, we aim here to study this case in detail and to show that the perturbative (cubic) description is insufficient in a wide range of configurations.

Specifically, we will solve Eq.~(\ref{eq:DT--polyn-norm}) and compute the permittivity~(\ref{eq:eps_T_linear}) and field~(\ref{eq:E_np}) for a varying level of illumination. Despite its considerable simplicity, the analytical solution of Eq.~(\ref{eq:DT--polyn-norm}) is too complicated for meaningful physical analysis. Accordingly, we proceed by a numerical solution of Eq.~(\ref{eq:DT--polyn-norm})%\footnote{We note that the 4-th order polynomial~(\ref{eq:DT--polyn-norm}) has only one physically meaningful solution. }
. In all the examples below, we will limit ourselves to a maximal temperature rise of $\Delta T \approx 300 ^\circ K \sim T_h^0$, a range in which the assumptions detailed above should hold~\cite{Shalaev_ellipsometry_gold}. In this range, one avoids sintering and melting of the metal~\cite{Munjiza_2014} as well as damage to the host. The quantitative discussion below will refer to the values obtained at the maximal incoming intensity.

We focus on Au and Ag, for which the temperature dependence of the permittivity was recently characterized~\cite{Wilson_deps_dT,Shalaev_ellipsometry_gold,PT_Shen_ellipsometry_gold,Indian_Ag_ellipsometry_2014}. We consider generic dielectric hosts such as an index matching oil and a (solid) semiconductor. Additional material choices can be analyzed along the same lines. Note we avoid water as a host in order to avoid complications associated with the limited range of temperature increase before boiling, super heating and bubble formation~\cite{Halas-bubble1,Halas-bubble2,Baffou-bubble1}. Our wavelength choices are rather customary, so that overall, the examples below are fairly generic.

\section{Numerical examples} \label{sec:examples}
\subsection{Au nanospheres in a liquid host}\label{subsec:Au_liquid}

We set $\lambda = 550$nm such that $\epsilon_{Au,0} = - 5.3 + 1.76 i$ and $B_{Au} = (0.7 + 1.7i)\cdot 10^{-3} / ^\circ K$~\cite{PT_Shen_ellipsometry_gold} so that with a liquid host for which $\epsilon_h = 2.65$, the illumination is resonant. We also set $\kappa_{h,0} = 0.15\ W/m/^\circ K$ and $B'_h = 10^{-4}/^\circ K$~\cite{oil_T_dependence}; these three parameters correspond to an index matching oil, used in the experiments of~\cite{plasmonic-SAX-ACS_phot,plasmonic-SAX-PRL,plasmonic-SAX-OE,plasmonic-SAX-rods-Ag,SUSI}. In the absence of concrete data, we estimate $B_{\kappa,h} = 5 \cdot 10^{-5}\ W/m/^\circ K^2$ such that $B_{\kappa,h} / \kappa_{h,0} \sim 3 \cdot 10^{-4}/^\circ K$, and at $\Delta T = 300 ^\circ K$, the change of $\kappa_h$ is $\sim 10\% $.

\begin{center}
\begin{figure}[h!]
  \centering{\includegraphics[width=17cm]{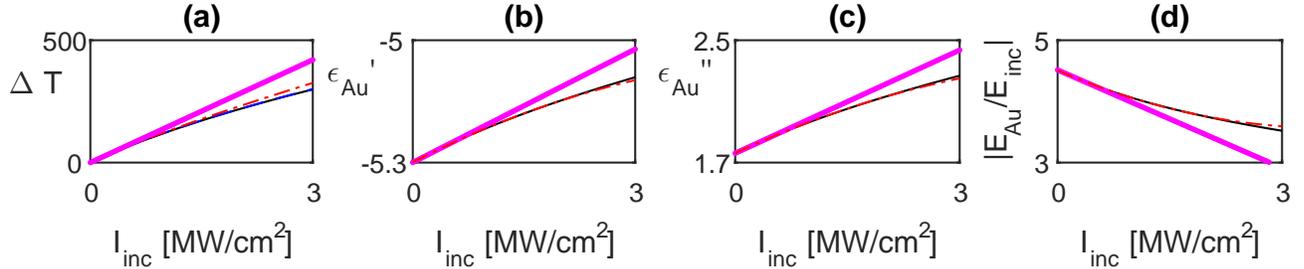}}
  \caption{(Color online) (a) The temperature rise (i.e., $\Delta T$, the solution of Eq.~(\ref{eq:DT--polyn-norm}); black solid line) inside a Au sphere in a liquid host ($\epsilon_h = 2.65$) illuminated at $\lambda = 550$nm as a function of the incident intensity. The solutions accounting for up to third/second/first order terms %{\bf denote them as $\Delta T^{I,II,III}$??}
  are shown by blue dashed/red dash-dotted/magenta dotted lines, respectively; however, in the current case, the black and blue lines are essentially indistinguishable. (b) Real and (c) imaginary parts of the Au permittivity as a function of the incoming intensity, based on the exact solution of Eq.~(\ref{eq:DT--polyn-norm}) (black solid line) compared with its cubic approximation (red dash-dotted line) and a cubic-quintic approximation (magenta dotted line). (d) Same as (b)-(c) for the normalized electric field within the nanosphere. }\label{fig:Au_oil_example}
\end{figure}
\end{center}

Fig.~\ref{fig:Au_oil_example}(a) shows that the temperature grows monotonically with the incoming intensity, but with a decreasing slope; we refer to this behaviour as a ``saturation'' of the thermal response. The deviation of the solution of the 4-th order polynomial~(\ref{eq:DT--polyn-norm}) from the standard linear solution~(\ref{eq:DT_lin_on}), i.e., the solution of Eq.~(\ref{eq:DT--polyn-norm}) with $a_2 = a_3 = a_4 = 0$, reaches several tens of percent. However, this deviation decreases to less than $10\%$ and even less than $0.5\%$ if the second order (i.e., with $a_2 \ne 0$) and third order ($a_2, a_3 \ne 0$) terms, respectively, are accounted for. The addition of the fourth-order term (i.e., also with $a_4 \ne 0$) has a very small effect.

Figs.~\ref{fig:Au_oil_example}(b)-(c) show the changes of the real and imaginary parts of the Au permittivity, respectively. The relative change of $\epsilon_{Au}'$ is a few percent, but the relative change of $\epsilon_{Au}''$ is $\sim 26\%$. Figs.~\ref{fig:Au_oil_example}(b)-(c) also show the fit to the exact solution of Eq.~(\ref{eq:DT--polyn-norm}), with either a first- or second-order polynomial in the incoming intensity, corresponding to a cubic or cubic-quintic approximation of the nonlinear response, respectively~\footnote{Accordingly, we adopt below the standard terminology of nonlinear optics, namely, the term proportional to $I_{inc} \sim \left|E_{inc}\right|^2$ will be referred to as the cubic term and the term proportional to $I_{inc}^2 \sim \left|E_{inc}\right|^4$ will be referred to as the quintic term.}. As seen, {\em the cubic fit provides a good match to the exact results only up to a temperature rise of a few tens of degrees}, so that the standard perturbative (cubic) description is inadequate. Instead, the cubic-quintic fit provides a decent match to the exact solution within the current temperature range, reflecting the ``saturation'' of the thermal effect, or equivalently, the decrease of the quality factor of the plasmonic resonator. % and is not very sensitive to the range over which the fit is performed.

Finally, Fig.~\ref{fig:Au_oil_example}(d) shows the field within the nanosphere. As expected, it decreases due to the increase of the imaginary part of the permittivity. Again, while the cubic fit is accurate only for a limited range of temperatures (hence, intensities), the cubic-quintic fit provides quite a satisfactory fit of the exact numerical results.

% {\bf study example with $\Delta \kappa_h$ negative and dominating! This will ACCELERATE the temperature growth rate! would this make it more nonlinear? - not so dramatic..}

Since $B_{tot}' \Delta T = O(10^{-1}) \ll \epsilon_{Au,0}''$ for the temperature rise of $O(100 ^\circ K)$ studied here, we see that typically, the changes of the field due to the changes of the real part of the Au permittivity, corresponding to a spectral shift of the resonance position, are smaller than those incurred by the changes to the imaginary part. In other words, the fact that the plasmonic resonance is typically broad, and the relative larger magnitude of $B_{Au}''$, make the system relatively less sensitive to changes of the real part of the Au permittivity. As shown below, this conclusion does not apply to Ag nanospheres. In that regard, a direct test of the importance of $B_m'$ and $B_h'$ shows that their neglect leads to a maximal error smaller than $1\%$ ($\sim 3 ^\circ K$) in the temperature rise (with respect to a $\sim 100\%$ total change of $\sim 300 ^\circ K$)~\footnote{Relative errors in the permittivity and field are of comparable magnitude. }. However, neglecting $B_{\kappa,h} / \kappa_{h,0}$ leads to a relative error of $\sim 7\%$ for the temperature rise~\footnote{i.e., a temperature increase of $323 ^\circ K$ instead of $300 ^\circ K$. The corresponding errors in the permittivity and field are similar. }. As shown below, this occurs because to leading order, $B_{\kappa,h} / \kappa_{h,0}$, affect the temperature rise on the same order as $B_m''/\epsilon_{m,0}''$, so that it is more influential compared to the other thermo-derivatives.

\subsection{Au nanospheres in a AlGaAs host} \label{subsec:Au_SC}

We now choose a solid host, specifically, Al$_x$Ga$_{1-x}$As with $x = 30\%$ Al content, such that the host is transparent up to the visible spectral range, and exhibits a rather high permittivity of $\sim 13$ near the band edge (i.e., about $5$ times higher than the oil liquid employed above). Accordingly, the localized plasmon resonance occurs at $\lambda = 800$nm where $\epsilon_{Au,0} = - 26.5 + 2 i$~\cite{Shalaev_ellipsometry_gold}; in addition, $B_{Au} = (2 + 2.6 i) \cdot 10^{-3} / ^\circ K$ which corresponds to the largest value of $B_{Au}''/\epsilon_{Au,0}''$ in the optical spectral domain. % sign of real part ambiguous..

This choice of Al content also corresponds to nearly minimal value of thermal conductivity, $\kappa_{h,0} \approx 13\ W/m/^\circ K$~\cite{ALGaAS-book}, i.e., about 100 times higher than for oil, but still about 20 times smaller than for metals. Thus, the error associated with neglecting the temperature non-uniformity amounts to no more than $2\%$~\cite{Sivan-Chu-high-T-nl-plasmonics}. This error is much smaller than all the effects demonstrated below.

% http://www.azom.com/article.aspx?ArticleID=8365;

Based on data for GaAs~\cite{permittivity_T_SCs}, we estimate % $B'_{AlGaAs} \approx 6 \cdot 10^{-3}/^\circ K$ (; (13-12.4) / 300K
$B'_{AlGaAs} \approx 2 \cdot 10^{-3}/^\circ K$%)
(such that $B'_{AlGaAs}/\epsilon_{AlGaAs,0}' \sim 10^{-4}/^\circ K$; this is similar to the liquid, and about an order of magnitude smaller than the corresponding ratio for metals~\footnote{As noted below, this is relation is fairly general, see Eq.~(\ref{eq:np_pp_dominance_h_p}).}) and estimate $B_{\kappa,h} = 5 \cdot 10^{-4} W/m/^\circ K^2$ such that at $\Delta T = 300 ^\circ K$, the change of $\kappa_h$ is $\sim 1\% $, i.e., it is negligible. We ignore strain/stress that may develop within the solid host due to the heating. These effects may only increase the nonlinear response. % {\bf .. but may dominate ...}

\begin{center}
\begin{figure}[h!]
  \centering{\includegraphics[width=17cm]{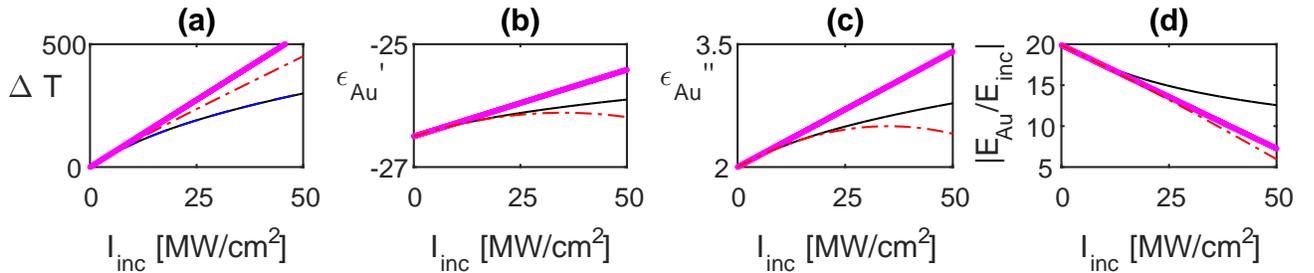}}
  \caption{(Color online) Same as Fig.~\ref{fig:Au_oil_example} for a Al$_{0.3}$Ga$_{0.7}$As host with $\lambda = 800$nm and $\epsilon_h = 13$. }\label{fig:Au_AlGaAs_example}
\end{figure}
\end{center}

Fig.~\ref{fig:Au_AlGaAs_example} shows results which are qualitatively similar to the case of liquid host (Fig.~\ref{fig:Au_oil_example}), however, for
%\footnote{but NOT more non-local... smaller $\kappa_h$ does not affect the temperature decay rate but the overall temperature rise; this conclusion is not unique to the quasi-static assumption. }
$\Delta T_{lin}^{on}$ which is 13 times smaller. This shows, in contrast to what one may expect based on the ultrafast case (see, e.g.,~\cite{West_alternat_plas_rev,NFE-thermo-plas-FOM}, Eqs.~(\ref{eq:E_np_explicit})-(\ref{eq:E_np_explicit_res}) and discussion below), that the metal-semiconductor system is {\em less} nonlinear than the metal-liquid system. This is a direct result of the heat diffusion in the host, and more generally, of the spatially non-local nature of the thermal nonlinearity in the system, causing lower effective heating of the metal nanosphere. % Notably, in the ultrafast case, the semiconductor system is expected to yield a far higher nonlinearity compared with the liquid system.

Another striking difference between Fig.~\ref{fig:Au_oil_example} and Fig.~\ref{fig:Au_AlGaAs_example} is that in the latter, neither the cubic nor the cubic-quintic fits provide a good match to the exact variations of the permittivity and field. As will be shown below, this can be traced to the relatively larger value of the (relative) change of the real part of the metal permittivity.

% {\bf quadratic approximation is not so good - T error is $> 100\%$.. $\epsilon_m''$ error is $> 40 \%$.. }

\subsection{Ag nanospheres in a liquid host}

We now replace the Au spheres with Ag spheres and set $\lambda = 450$nm such that $\epsilon_{Ag,0} = -4.7 + 1.55i$ and~$B_{Ag} = (-46 + 2.5 i) \cdot 10^{-4}/ ^\circ K$~\cite{Indian_Ag_ellipsometry_2014}. Note that at this wavelength, the thermo-derivative of the real part of the metal permittivity is 7 times higher than Au, and has the opposite sign; in addition, the thermo-derivative of the imaginary part is much smaller than in previous examples. % This can be related to the dominance of thermal expansion

% however, we should note that the parameters are sensitive to the wavelength and to the quality of the ellipsometry data.

% \footnote{\bf We should be careful with the data used here - based on a single work... add/mention Shalaev's data.. } % for lambda = 381nm, $\epsilon_{Ag,0} = -2.5 + 1.08i$ and~$B_{Ag} = (-0.57 + 1.7 i) \cdot 10^{-5}/ ^\circ K$, i.e., less sensitive to temperature.. + I do not know what would be the appropriate kappa for the host (sol-gel??) for 354nm, the real part of the Ag permittivity is already too low..

\begin{center}
\begin{figure}[h!]
  \centering{\includegraphics[width=17cm]{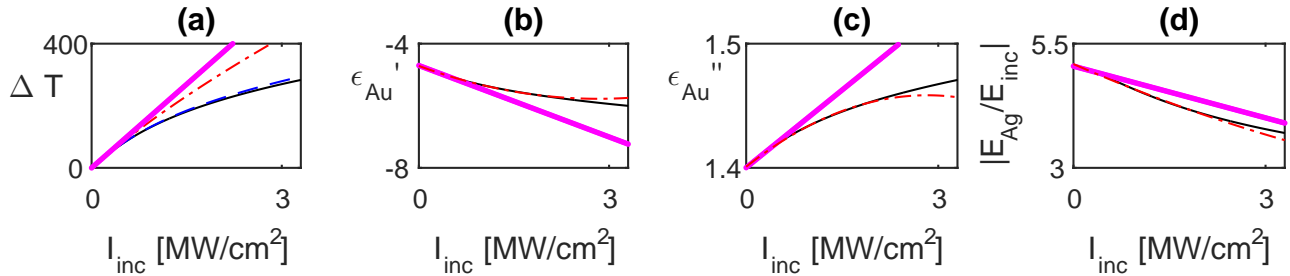}}
  \caption{(Color online) Same as Fig.~\ref{fig:Au_oil_example} with Ag nanospheres and $\lambda = 450$nm and $\epsilon_h = 2.35$. }\label{fig:Ag_oil_example}
\end{figure}
\end{center}

The temperature rise shown in Fig.~\ref{fig:Ag_oil_example}(a) is reminiscent to those shown above, however, in contrast to previous cases, the real part of the permittivity of the Ag particle decreases (rather than increases). More importantly, although Fig.~\ref{fig:Ag_oil_example}(d) shows results which look quantitatively similar to the cases of Au particles, the reason for the decrease of scattering for the Ag particles is different - it originates from the change of the real part of the permittivity rather than the change of the imaginary part, i.e., the field decreases due to a shift {\em away from resonance}, rather than due to a decrease of the quality factor of the resonator. Indeed, the change of the real part (Fig.~\ref{fig:Ag_oil_example}(b)) is far larger than the change of the imaginary part (Fig.~\ref{fig:Ag_oil_example}(c)). This dominance also makes the cubic-quintic fit quite satisfactory, in contrast to what is seen in Fig.~\ref{fig:Au_AlGaAs_example}.

\begin{center}
\begin{figure}[h!]
  \centering{\includegraphics[width=17cm]{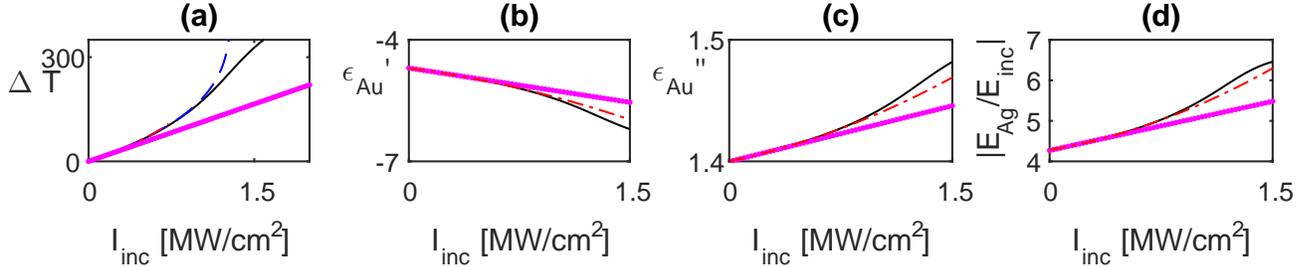}}
  \caption{(Color online) Same as Fig.~\ref{fig:Ag_oil_example} with $\epsilon_h = 3.25$. }\label{fig:Ag_oil_example2}
\end{figure}
\end{center}

This observation prompts us to predict the opposite behaviour, so far not observed experimentally. Specifically, if the illumination is initially non-resonant, the heat generation can cause the system to shift {\em into} resonance, hence, the quality factor and local field amplitude to increase. This is shown in Fig.~\ref{fig:Ag_oil_example2}. Peculiarly, this example shows a deviation from the perturbative description in (a), yet, (b)-(d) are well described by a cubic-quintic nonlinearity.

% Similar behaviour is observed for a semiconductor host.

% the maximal Ag enhancement I see is much less than Baffou's 118, because I do not work at resonance, where the imaginary part is smallest.

% \footnote{The experimental data shows that the thermo-derivative at 354nm, where $\epsilon_{Ag}''$ is minimal, is much smaller BY HOW MUCH, so that the nonlinear response is, in fact, smaller than for the set of parameters chosen..}

\section{More analysis and simplification} \label{sec:analysis}

We now try to simplify Eq.~(\ref{eq:DT--polyn-norm}) by identifying the leading order terms, in order to complement the numerical solutions shown above with an approximate analysis, valid for the moderately high intensity range. This is especially important in order to emphasize the deviation from the perturbative description of the nonlinear response and to elucidate the relative importance of the various physical effects and how each of them affects the temperature, permittivity and field.

The easiest way to obtain analytical insights is, as already done in the numerical simulations above (see red lines in Figs.~\ref{fig:Au_oil_example}(a) \&~\ref{fig:Au_AlGaAs_example}(a)), to limit ourselves to small intensities, hence, to a small temperature rise, such that it is justified to keep only the terms up to the second order of $\Delta T$ in Eq.~(\ref{eq:DT--polyn-norm}), i.e., to set $a_3 = a_4 = 0$. In this case,
\begin{eqnarray}\label{eq:DT_II}
\Delta T &=& % \frac{- a_1 + \sqrt{{a_1}^2 + 4 a_2 \Delta T_{lin}^{on}}}{2 a_2} =
\frac{a_1}{2 a_2} \left(\sqrt{1 + 4 \frac{a_2}{{a_1}^2} \Delta T_{lin}^{on}} - 1\right).
\end{eqnarray}
This shows that in general, % for sufficiently high temperatures, i.e., for temperature rise of $100-200 ^\circ K$ or even more,
the temperature rise does not scale linearly with the incoming intensity (via $\Delta T_{lin}$~(\ref{eq:DT_lin})) as one would expect from the perturbative description; instead, it follows a square-root law, i.e., it grows more slowly (for $a_2 > 0$, which is the usual case, see discussion below); we referred to this behaviour as a ``saturation'' of the thermal effect. This is in agreement with the numerical results shown above. % \footnote{who showed a decrease of the normalized absorption}

For even smaller intensities, we can expand the square root in a Taylor series~\footnote{Note that the expansion~(\ref{eq:DT_II_Taylor}) is with respect to the {\em quadratic} solution, i.e., it is less accurate than the cubic-quintic fits to the {\em exact} solution shown in the plots of the nonlinear permittivities. Yet, in most cases, this gives a decent guidance regarding the relative importance of the various terms. } such that
%\begin{equation}\label{eq:Delta_T_gen_res}
%\Delta T  = \frac{a_1}{4 a_2} \left(\Delta - \frac{\Delta^2}{4} + \frac{\Delta^3}{8} - \cdots \right),
%\end{equation}
%where
%\begin{equation}\label{}
%\Delta \equiv 4 \frac{a_2}{{a_1}^2} \Delta T_{lin}^{on}.
%\end{equation}
%Substituting gives
\begin{eqnarray}\label{eq:DT_II_Taylor}
\Delta T &=& % \frac{a_1}{4 a_2} \left(4 \frac{a_2}{a_1^2} \Delta T_{lin}^{on} - \frac{\left(4 \frac{a_2}{a_1^2} \Delta T_{lin}^{on}\right)^2}{4} + \cdots \right) =
\frac{\Delta T_{lin}^{on}}{{a_1}} - \frac{a_2}{a_1^3} \left(\Delta T_{lin}^{on}\right)^2 + \cdots, % = \frac{1}{a_1}\frac{3 a^2 \omega n_{h,0}^3 I_{inc}}{4 c \kappa_{h,0} \epsilon''_{m,0}} - \frac{a_2}{a_1^3} \left(\frac{3 a^2 \omega n_{h,0}^3 I_{inc}}{4 c \kappa_{h,0} \epsilon''_{m,0}}\right)^2 + \cdots.
\end{eqnarray}
where $\Delta T_{lin}^{on}$~(\ref{eq:DT_lin_on}) scales with $I_{inc}$, $a^2$, and $\kappa_{h,0}^{-1}$; similarly, the nonlinear correction to the temperature scales with $I^2_{inc}$, $a^4$, and $\kappa_{h,0}^{-2}$.

Eqs.~(\ref{eq:eps_T_linear}), (\ref{eq:DT--polyn-norm}) and~(\ref{eq:DT_II_Taylor}) show that the relative importance of a specific physical effect (i.e., a specific thermo-derivative) is related to the highest order coefficient of~(\ref{eq:DT--polyn-norm}) it appears in, e.g., the cubic term depends on $a_1$, the quintic depends also on $a_2$ etc.. Specifically, this shows that $B_{\kappa,h} / \kappa_{h,0}$ affects the temperature rise and field~(\ref{eq:E_np_explicit}) only via the quintic term.

We also note, however, that some of the contributions to the coefficients $a_1$ to $a_3$ are proportional to $\Delta T_{lin}^{on}$. These will, naturally, contribute only to the next order in $\Delta T_{lin}^{on}$ (or equivalently, in $I_{inc}$). % thus, it is not justified to neglect them!
Specifically, this shows that although $B_h'$ appears in $a_1$, it too affects the temperature rise $\Delta T$ only via the quintic term. Note that somewhat unintuitively, this is lower than its effect on the field~(\ref{eq:E_np}).

Further distinction between the various contributions can be attained by noting that typically, the permittivity of a (host) {\em dielectric} material is less sensitive to temperature than the metal, i.e.,
\begin{equation}\label{eq:np_pp_dominance_h_p}
B_h'/\epsilon_{h,0}' \ll \left|B_m'/\epsilon_{m,0}''\right|,\ \left|B_m'/\epsilon_{m,0}'\right|,\ \left|B_m''/\epsilon_{m,0}''\right|.
\end{equation}
Indeed, one can compare e.g., recent ellipsometry results~\cite{Shalaev_ellipsometry_gold,PT_Shen_ellipsometry_gold,Indian_Ag_ellipsometry_2014} with values detailed in~\cite{Boyd-book}; see also specific examples above~\footnote{Eq.~(\ref{eq:np_pp_dominance_h_p}) enables one to simplify some of the coefficients in Eq.~(\ref{eq:DT--polyn-norm}). However, as shown below regarding the quadratic approximation~(\ref{eq:DT_II}), this approach has only a limited accuracy. }.
% Because of these reasoning, $a_2 \sim B_{m,0}'' > 0$ for most of the spectrum, see~\cite{Indian_Ag_ellipsometry_2014,PT_Shen_ellipsometry_gold,Shalaev_ellipsometry_gold}; - ACTUALLY - NIT TRUE!
Accordingly, although it appears in the quintic term, the contribution of $B_h'$ to the temperature rise $\Delta T$ is small relative to $B_m''$ such that it affects $\Delta T$ only for very high incoming intensities, beyond the cubic-quintic description. % Similarly, we will see below that $B_m'$ affects $\Delta T$ only beyond the quintic order.
Again, this is lower than its effect on the field~(\ref{eq:E_np}).

% As we shall see below, the relative importance of the other ???????? thermo-derivatives varies depending on
Even further distinction can be made for a specific spectral configuration. At resonance, i.e., for $\epsilon_{m,0}'' \gg \epsilon'_{tot} \to 0$, Eq.~(\ref{eq:DT--polyn-norm}) reduces to
\begin{eqnarray}\label{eq:T_2nd_orderG}\label{eq:DT--polyn-norm-approx5}
\underbrace{\left[2 \frac{B_m''}{\epsilon_{m,0}''} + \frac{B_{\kappa,h}}{\kappa_{h,0}} - \Delta T_{lin}^{on} \left(2\frac{B''_m B'_h}{\epsilon_{m,0}'' \epsilon_{h,0}'} + \frac{{B'}^2_h}{\epsilon_{h,0}'^2}\right)\right]}_{a_2} \Delta T^2 + \underbrace{\left[1 - \Delta T_{lin}^{on} \left(\frac{B_m''}{\epsilon_{m,0}''} + 2 \frac{B'_h}{\epsilon_{h,0}'}\right)\right]}_{a_1} \Delta T  = \Delta T_{lin}^{on}. \nn \\
\end{eqnarray}
% This equation should be in principle more accurate than Eq.~(\ref{eq:DT--polyn-norm-approx4}), as it accounts for $B_{\kappa,h} / \kappa_h$?? and $B_h'$..
Organizing the orders of $\Delta T_{lin}^{on}$, Eq.~(\ref{eq:DT_II_Taylor}) reduces to
\begin{eqnarray}\label{eq:DT_II_Taylor_on}
\Delta T &=& \Delta T_{lin}^{on} - \left(\frac{B_m''}{\epsilon_{m,0}''} + \frac{B_{\kappa,h}}{\kappa_{h,0}} + 2 \frac{B_h'}{\epsilon_{h,0}'}\right) \left(\Delta T_{lin}^{on}\right)^2 + \cdots.
\end{eqnarray}

Thus, the thermo-derivatives affect the temperature rise $\Delta T$ via the quintic or higher order terms. % Specifically, the quintic response relies on $B_m''/\epsilon_{m,0}''$, $B_{\kappa,h}/\kappa_{h,0}$ and $B_h'/\epsilon_{h,0}'$.
Note that the various terms may not necessarily have the same sign, so that they can balance each other. However, by Eq.~(\ref{eq:np_pp_dominance_h_p}), $B_h'/\epsilon_{h,0}'$, is typically smaller with respect to $B_m''/\epsilon_{m,0}''$, so that the sign of $B_h'$ is of lesser importance and the quintic coefficient can be simplified. In addition, Eq.~(\ref{eq:DT_II_Taylor_on}) shows that in this case, $B_m'$ appears only beyond the quintic-order. These two facts a-posteriory justify the numerical results shown in~\cite{Sivan-Chu-high-T-nl-plasmonics} and in Section~\ref{subsec:Au_liquid}. % . where $B_m'/\epsilon_{m,0}''$ was shown to have a relatively small role.

In addition, Eq.~(\ref{eq:DT_II_Taylor_on}) explains the failure of the cubic-quintic fit in the Au-AlGaAs case (Section~\ref{subsec:Au_SC} and Fig.~\ref{fig:Au_AlGaAs_example}). Indeed, in this case, the relative larger value taken by $B_{Au}'$ makes the $a_3$ term in Eq.~(\ref{eq:DT--polyn-norm}) relatively more important than in the liquid host case, hence, it gives rise to a deviation from the cubic-(quintic) description.

However, Eq.~(\ref{eq:DT_II_Taylor_on}), which does not account for $B_m'$, fails to reproduce the results of Figs.~\ref{fig:Ag_oil_example}-\ref{fig:Ag_oil_example2} where the contributions of $B_{\kappa,h}$ and $B_m''$ are relatively small. Indeed, in these cases, the $a_3$ and $a_4$ terms of Eq.~(\ref{eq:DT--polyn-norm}) are of comparable magnitude to that of the $a_2$ term already for a small temperature rise and the approximation~(\ref{eq:DT_II}) is not valid anymore.

Away from resonance, i.e., for $\epsilon'_{tot} \gg \epsilon_{m,0}''$, Eq.~(\ref{eq:DT_II_Taylor}) can be approximated by
\begin{eqnarray}\label{eq:DT_II_Taylor_off}
\Delta T &\approx& \left(\frac{\epsilon_{m,0}''}{\epsilon'_{tot}}\right)^2 \Delta T_{lin}^{on} - \left(\frac{B_m''}{\epsilon_{m,0}''} + 2 \frac{B'_h}{\epsilon_{h,0}'} + \left[2 \frac{B'_{tot} \epsilon'_{tot}}{{\epsilon_{m,0}''}^2} + \frac{B_{\kappa,h}}{\kappa_{h,0}} \left(\frac{\epsilon_{m,0}''}{\epsilon'_{tot}}\right)^{-2} \right] \left( \frac{\epsilon_{m,0}''}{\epsilon'_{tot}}\right)^6\right) \left(\Delta T_{lin}^{on}\right)^2 + \cdots. \nn \\
% a_1 &=& \left[\left(\frac{\epsilon'_{tot}}{\epsilon_{m,0}''}\right)^2 - \Delta T_{lin}^{on} \left(\frac{B_m''}{\epsilon_{m,0}''} + 2 \frac{B'_h}{\epsilon_{h,0}'}\right)\right] \nn \\
% a_2 &=& \left[2 \frac{B'_{tot} \epsilon'_{tot} + B_m'' \epsilon''_{m,0}}{{\epsilon_{m,0}''}^2} + \frac{B_{\kappa,h}}{\kappa_{h,0}} \left(\frac{\epsilon'_{tot}}{\epsilon_{m,0}''}\right)^2 \right]
\end{eqnarray}
Thus, with respect to the on-resonance case, the cubic term is smaller and for the quintic case, the weights of the various thermo-derivatives are different. Specifically, the quintic term now includes also $B_m'$. % Still, by Eq.~(\ref{eq:np_pp_dominance_h_p}), $B_m''$ is the dominant thermo-derivative.
This explains the relatively larger role played by $B_{Ag}'$ in Figs.~\ref{fig:Ag_oil_example} and~\ref{fig:Ag_oil_example2} compared to all other examples. Eq.~(\ref{eq:DT_II_Taylor_off}) would be also relevant to thin films (assuming no means for phase matching is adopted).

% {\bf red line invalid... }

\section{Discussion} \label{sec:discussion}

% \subsection{Description of results}

The numerical examples and consequent approximate analysis above show that the nonlinear thermal response is cubic only for a modest temperature rise of a few tens of degrees; this is a regime where the nonlinear response is, in fact, frequently ignored altogether~\cite{thermo-plasmonics-review}. Beyond this range, a cubic-quintic nonlinear response is sufficient in cases where either $B_m'$ or $B_m''$ dominate all other thermo-derivatives. Otherwise, a fully non-perturbative description of the thermal response is required.

Overall, the changes of the metal permittivity dominate the nonlinear thermal response of the nanosphere configuration. However, this does not mean that the effects of thermal lensing due to heating of the host are secondary. Indeed, the host occupies, in general, a much larger fraction of space; their role should anyhow be studied in the context of effective medium theory, which is beyond the scope of the current study.

We note that all the cases studied here involved $B_m'' > 0$, leading typically to the ``thermal saturation'' described above - the temperature grows more slowly at high intensities compared to low intensities. In turn, this leads to a decrease of the internal and scattered fields with increasing intensity~\cite{Sivan-Chu-high-T-nl-plasmonics}. This behaviour was indeed observed in all experiments of nonlinear scattering from metal nanoparticles under CW illumination performed so far~\cite{plasmonic-SAX-ACS_phot,plasmonic-SAX-PRL,plasmonic-SAX-OE,plasmonic-SAX-rods-Ag,SUSI}\footnote{As an aside, we note that a similar behaviour was seen in~\cite[Fig.~1]{Hache-cubic-metal-nlty} for dilute composites under {\em picosecond} illumination. However, there are also reports of a potentially opposite behaviour (e.g.,~\cite{Gurudas_JAP_2008,Elim_APL_2006,IOSB}). This reflects the sensitivity of the problem to the illumination wavelength, scatterer size and for pulsed illumination, also for the duration of the illumination. }. However, the ellipsometry data reveals narrow spectral regimes where $B_m'' < 0$, e.g., for Au for $\lambda < 500nm$ (due to the effect of the Fermi smearing on the interband transitions~\cite{Rosei_Au_diel_function,Rosei_Au_diel_function2}). This regime is, however, out of the resonance band, hence, the response of the system is expected to be weak in this regime. For Ag~\cite{Indian_Ag_ellipsometry_2014}, a much narrower regime in which $B_m'' < 0$ around $\lambda = 400$nm was identified. Even then, the thermo-derivatives in this regime are much smaller than in the examples above. Yet, although elusive, such a regime is clearly highly desirable for plasmonic applications, since it gives rise to higher quality factors and {\em stronger} fields. %... but not sure will survive for small NPs ...
We emphasize, however, that this effect should not be confused with saturable absorption (= partial population inversion, thus, associated with a non-thermal electron distribution), which could explain the intriguing ``reverse saturable'' scattering observed in~\cite{plasmonic-SAX-ACS_phot,plasmonic-SAX-PRL,plasmonic-SAX-OE,plasmonic-SAX-rods-Ag}\footnote{Such increased scattering was observed for ultrafast illumination, see e.g.~\cite{Gurudas_JAP_2008,Elim_APL_2006,IOSB}, and was repeatedly interpreted as a saturable absorption; however, to the best of our knowledge, no conclusive proof regarding the origin of the effect was ever given.}.

% \subsection{Physics}

% \subsection{Figure of Merit}

It is tempting to write the temperature dependent term in Eq.~(\ref{eq:eps_T_linear}) as the product of the intensity and the cubic nonlinear coefficient associated with the thermal effect, and similarly for the quintic effect etc.. % Indeed, similar to Boyd, but for more than one material..%  this is implied in many studies (SWC vs. Iranians.. ) and used to compare the thermal nonlinearity to other nonlinearities. ...
However, strictly speaking, such as expansion~(\ref{eq:eps_T_linear}) {\em cannot} be regarded as an inherent material parameter, because although it involves the intrinsic material electromagnetic response of the metal, it also involves, as would happen for any structure, the size of the nanosphere, and the electromagnetic and thermal response of the host\footnote{We expect the shape to affect the cubic coefficient as well, however, a quantification of this effect requires more exact numerical calculations that we will perform only in a future study. }. % \footnote{Something similar happens in the ultrafast case, where $\epsilon_d$ plays a role in determining the field within the metal {\bf this will always be the case.. no?}. This required a careful separation of the effective and intrinsic metal nonlinearity, see e.g.,~\cite{Hache-cubic-metal-nlty,Hamanaka-Nakamura-2003,Nakamura-96,Liao_OL_1998,Smith_Boyd_EMT_kerr_media,Smith_Boyd_EMT_kerr_media_1999}. }; this is an obvious but sometimes overlooked fact.
On the other hand, the temperature-dependent term~(\ref{eq:eps_T_linear}) cannot be understood as an effective nonlinearity either, since such a quantity necessarily involves the fill factor, inter-particle interactions etc..

Thus, the correction to the permittivity, $B_m \Delta T$, or its leading order term $B_m \Delta T_{lin}$, should be understood as a Figure of Merit (FOM) for the nonlinear response. Specifically, the values of nonlinear coefficients reported in~\cite{SUSI} cannot be compared to other reported values of intrinsic nonlinearity of metals.

% {\bf assuming the thermal effect dominated the Kerr effect...} not only the analysis implies on the incorrect effect (pure cubic local response rather than a generalized nonlinear nonlocal response).

Note that our FOM, $B_m \Delta T_{lin}$, is proportional to the absorption cross-section of the nanosphere or equivalently, to the Joule number~\cite{NFE-thermo-plas-FOM}. It, however, includes also information about the absolute temperature rise for a given incoming intensity and host by accounting for heat diffusion in the host. Accordingly, while the conventional FOM (Eq.~(\ref{eq:f}) and~\cite{West_alternat_plas_rev,NFE-thermo-plas-FOM}) is relevant globally for linear optics, in the context of intense illumination, it is useful only for ultrafast illumination where the thermal properties do not play an important role, i.e., when the response is spatially-local. In all other cases, especially the many associated applications, our FOM should be used. Specifically, our FOM shows that the highest thermal nonlinearity will be obtained for a dilute host (e.g., gas, or even vacuum), due to its low thermal conductivity.

% This has to be taken into account for the relevant applications.. % As a result, our FOM is relevant for applications based on intense CW illumination, of high temperature plasmonics compared to the standard FOM.
% he deals with a fixed host to compare different metals, but I am talking about comparing different hosts...

% Baffou FOM~\cite{NFE-thermo-plas-FOM} - only plasmonic/EM aspects, hence, of general nature. distinguishes between FOM for scattering and for absorption.  - we specialize their results for a thermal nonlinearity + quantify the temperature rise and show the importance of going beyond the linear approximation ...

% {\bf Still, in order to enable a fair comparison between different choices of materials(?), one can assume the same values for the dielectric host. his quantity indicates tendency to get heated.. ours, $\Delta T_{lin}^{on}$, quantifies the actual temperature rise, as well as the change to the permittivity and field (???).. or need $B_m('')$??}

% {\bf But this is smaller than liquid crystal nlties - $\sim 10^{-3} cm^2/W$ - 1000 times more than here! but maybe because such temperature rises cannot be achieved with LCs + LC devices not typically associated with strong focusing, then such nlties were not demonstrated before..? maybe consult Ibrahim? what is the damage threshold of LC's for CW illumination? }

The results of the current manuscript are specific to a few nm sphere. However, we believe that our claims apply also to larger spheres, as well as to particles of other shapes, with only quantitative differences associated with the non-uniformity of the field and temperature. % for larger/different NPs effects will be similar (but stronger/weaker? I expect the effective nlty to decrease for larger NPs, since the heating is not as effective..)

The results presented above and the good match to experimental results~\cite{Sivan-Chu-high-T-nl-plasmonics} support the interpretation of the experimental results of nonlinear scattering of CW light from the small nanospheres as a thermal nonlinear effect. However, many effects which do not play an important role in the ultrafast case, may have a non-negligible contribution in the CW case; these include specifically, stress/strain, partial population inversion due to interband transitions~\cite{Hache-cubic-metal-nlty}, free-carrier generation, multi-photon absorption etc.. In that regard, our study identifies several easy measurements that may allow us to establish the dominance of the thermal effect over all the additional ones.

First, we show that the variation of $\kappa_h$, which appears only in the thermo-optical nonlinearity studied here, may enable the isolation experimentally the role of the thermal effect. An attempt to do that was made in~\cite{Iranians_kappa_nlty}, however, this study ignored various additional relevant factors like the size of the particles that affects the leading order solution $\Delta T_{lin}$~(\ref{eq:DT_lin}). It also compared composites with different densities (details not provided) that could also affect the effective response due to the long range nature of the thermal interactions between adjacent nanospheres. These facts can explain the large discrepancy in~\cite{Iranians_kappa_nlty} between the theoretical prediction (based solely on the value of the thermal conductivity) and the experimental observations. A more accurate match requires accounting for all the additional parameters, including $B_{\kappa,h}$ which appears on the quintic order~(\ref{eq:DT_II_Taylor_on}), for size effects by relaxing the assumptions on field and temperature uniformity, as well as for the potential absorption in the host. Alternatively, one has to compare the measurement of a composite to a full effective medium theory for a composite. % , a theory we will develop in a future publication.

Second, the configuration described in Fig.~\ref{fig:Ag_oil_example2} is related, again, with a unique signature of the thermal effect. However, we emphasize that there is still some uncertainty regarding the exact values of $B_m$ for Ag, especially due possible differences of morphology of particles made with different methods (see e.g., discussion in~\cite{PT_Shen_ellipsometry_gold}). % This issue requires further study, potentially even require the use of alternative ways to determine the temperature dependence of the Ag permittivity, such as in~\cite{Boyd-SPR-eps-T}.

Finally, although the most accessible experimental indication for nonlinear scattering is the far field intensity, there is growing interest in measuring temperature in the near field~\cite{Pramod_Reddy_T_measure_2010,Lukin_T_measure,Mecklenburg_T_measure,NTU_T_measure_2015,Pramod_Reddy_T_measure_2015,Baumberg_SERS_T_measure}. Such measurements will give the ultimate evidence regarding the role of the thermal effect. The formulation, numerical results and analysis described in this manuscript will provide valuable predictions for such measurements.

% \bigskip

% \bigskip

In this study, we limited ourselves the thermal effects and modest temperature increases. Beyond that range, the thermo-derivatives are not necessarily constant, giving rise to further deviation from the perturbative description. In addition, other effects may become important for higher temperatures/intensities, e.g., multi-photon absorption, and non-thermal effects~\footnote{the latter effect may explain the rather surprising increase of scattering observed for high intensities that seem to dominate the thermal effect~\cite{plasmonic-SAX-ACS_phot,plasmonic-SAX-PRL,plasmonic-SAX-OE,plasmonic-SAX-rods-Ag}.}. In the presence of the latter, one would not be able to use the high temperature ellipsometry results of~\cite{Shalaev_ellipsometry_gold,PT_Shen_ellipsometry_gold} which are relevant only for metals are at thermal equilibrium, i.e., their electrons obey the Fermi-Dirac distribution. Clearly, at even higher temperatures, the metal would undergo sintering and eventually melting, and the host may get damaged.

As a final statement, this study provides an important step towards the characterization of the effective response of arrays of particles, which can be regarded as thermo-optical metamaterials - artificial materials that change their electromagnetic properties under laser illumination due to temperature rise. These would have to be studied under proper effective medium theory and beyond the perturbative description, accounting not only for the electromagnetic interactions between the nanoparticles, but also for the thermal interaction between them. The latter is expected to have a far longer length scale. This fact, together with the use of an absorptive nonlinearity, may allow surpassing the limitations detailed in~\cite{Khurgin_chi3} for highly nonlinear metamaterials.

\bibliographystyle{unsrt}
% \bibliography{C:/Users/Yonatan/Documents/Research/my_bib}
% \bibliography{D:/MyDocs/Research/my_bib}

\end{document}